\documentclass[smallextended,english]{svjour3}     
\usepackage[utf8]{inputenc}
\usepackage[english]{babel}
 
\usepackage{color}
\usepackage{graphicx}
\usepackage{amsmath}
\usepackage{amssymb}
\usepackage[round]{natbib}
\usepackage{bm}
\usepackage{bbm}
\usepackage{sidecap}

 \usepackage[T1]{fontenc}

\smartqed 

\begin{document}

\title{PARAMETER INFERENCE FROM HITTING TIMES FOR PERTURBED BROWNIAN
  MOTION} 
\titlerunning{Parameter inference from hitting times}

\author{M. Tamborrino\and S. Ditlevsen\and\\ P. Lansky}
\authorrunning{Tamborrino, Ditlevsen, Lansky}

\institute{P. Lansky \at Institute of Physiology, Academy of Sciences of the Czech Republic,Videnska 1083,\\
Prague 4, 142 20, Czech Republic\\
 \email{lansky@biomed.cas.cz}
\and M.Tamborrino \and S. Ditlevsen \at Department of Mathematical Sciences, Copenhagen University, Universitetsparken 5, DK-2100, Copenhagen, Denmark \\
Tel.: +45 35320785\\
Fax: +45 35320704 \\
\email{mt@math.ku.dk, susanne@math.ku.dk}}
\date{}
\maketitle

\begin{abstract}
A latent internal process describes the state of some system, e.g. the social tension in a political conflict, the strength of an industrial component or the health status of a person. When this process reaches a predefined threshold, the process terminates and an observable event occurs, e.g. the political conflict finishes, the industrial component breaks down or the person has a heart attack.  Imagine an intervention, e.g., a political decision, maintenance of a component or a medical treatment, is initiated to the process before the event occurs. How can we evaluate whether the intervention had an effect?

To answer this question we describe the effect of the intervention through parameter changes of the law governing the internal process. Then, the time interval between the start of the process and the final
event is divided into two subintervals: the time from the start to the
instant of intervention, denoted by $S$, and the time between the intervention and the threshold crossing,
denoted by $R$.  The first question studied here is: What is the joint distribution of $(S,R)$? The theoretical expression is provided and serves as a basis to answer the main question: Can we estimate the parameters of the model from observations of $S$ and $R$ and compare them statistically? Maximum likelihood estimators are illustrated on simulated data under the assumption that the process before and after the intervention is described by the same type of model, i.e. a Brownian motion, but with different parameters.
\keywords{first passage times; maximum likelihood estimation; Wiener process; degradation process; reliability; effect of intervention}
\end{abstract}

\section{Introduction}
Statistical inference for univariate stochastic processes from observations of hitting times, i.e. epochs when the process attains a boundary for the first time, is a common problem, see \cite{LeeWhitmore2006} and references therein. Here we investigate its specific variant for perturbed stochastic processes and discuss it in a general setting, presenting some of the fields in which this methodology can be applied. 
At a known time instant, either controlled by an experimentalist  or induced by an independent external condition, an intervention is initiated and the time to a given event following the intervention is measured. Assume that the intervention causes a change in the parameters of the underlying process. This scenario can be found in many fields, such as reliability theory, social sciences, finance, biology or medicine. The time course of the intervention can be interpreted as a time-varying explanatory factor in a threshold regression.

For analysing reliability of technical systems it is important to investigate damage processes. A common model is the Wiener process 
\citep{Whitmore1995,Whitmore1997,Whitmore1998,Whitmore2012,Kahle1998}.
In \cite{Pieper1997}, changing drifts of Wiener processes describes various stress levels for a damage process. 
\cite{DoksumHoyland1992} use a Gaussian process
and inverse Gaussian distribution (IGD) to discuss a lifetime model under a
step-stress accelerated life test. \cite{Nelson2008} discusses practical issues when conducting an
accelerated life test. \cite{Yu2003} proposed a systematic approach to the classification problem
where the products' degradation paths satisfy Wiener processes. Our model fits into the above framework as follows. The degradation of a component is modeled by a Wiener process with failure corresponding to the first crossing of a certain level. The time for maintenance is independent of the time since last repair and the maintenance changes the parameters of the Wiener process. Then from measurements of the time from last repair to
the time of maintenance and from the maintenance to the degradation, we deduce the effect of the maintenance on the system. Similarly to technical systems, a 
degradation process in a medical context is commonly modeled as an intrinsic, but not observable, diffusion stochastic process. With this interpretation, our 
model takes into account an abrupt change of
medication or life style before an observable event takes place. For example, in  \cite{CommengesHejblum2013} the event is myocardial infarction or coronary heart disease and the degradation is the atheromatous process, which is modeled as a Brownian motion with drift, where the drift is a function of explanatory variables. 

\cite{Lancaster1972} makes effective use of the IGD in describing data on 
duration of strikes in UK between 1965 and 1972. The rationale  is based on the idea of an underlying Wiener process. Despite that
alternative distributions are proposed
\citep{Kennan1985,Lawrence1984,NewbyWinterton1983}, the approach via
the first passage time (FPT) of the 
Wiener process remains one of the alternatives \citep{HarrisonStewart1993,DesmondYang2011}. \cite{Linden2000} extends 
Lancaster's approach by deriving the strike duration density from a controlled Wiener process. The FPT distribution of a controlled Wiener process is related to IGD, and it is shown  that since the maximum-likelihood estimates of expected strike duration with FPT density from a controlled process and IGD are the same, the IGD case offers a simple and valid approach to the analysis of the strike duration. Again, the model studied in this paper fits this reality. Imagine that during a strike an important offer towards strikers is proposed. Then the time after may move on a different scale.

In neuroscience, the interval between two consecutive action potentials is often studied being related to information transfer in neurons. The Wiener process is sometimes chosen to model the subthreshold membrane potential evolution of the neuron \citep{GersteinMandelbrot} and parameter estimation has been investigated \citep{ReviewPetr}. Moreover, estimation from observations of the last action potential before the intervention and the next following it, also in presence of delayed response to the stimulus, has been investigated \citep{TDL, TDL2}. The current model also fits this framework.

The aim of this paper is to solve two problems. The first  is the investigation of the joint distribution of the subintervals up to the instant of intervention, and between the intervention and the first crossing after it. This is needed for the second problem, namely the estimation of the parameters of the process before and after the intervention and testing their equality. Obviously, the two subintervals are dependent and the statistical inference is complicated by not observing the position of the process at the time of intervention. The main contributions of the paper are the solutions to these questions in the case of a perturbed Brownian motion. A detailed guideline on how to carry out both simulation of the data  and parameter estimation in the computing environment \textbf{R}  \citep{R}  is presented (see Appendices \ref{Appendix3} and \ref{Appendix2}).  Using the derived theoretical expressions, estimation could be carried out for more complicated diffusion processes.

In Section \ref{Section2} the character of experimental data together with a description of the involved quantities and variables are presented. In Section \ref{Section3} we  describe the model, mathematically define the quantities of interest and derive the probability densities for a general diffusion process. The Brownian motion model under different assumptions on its parameters is treated in Section \ref{secWiener}. The performance of maximum likelihood estimators and testing the difference between parameters are illustrated in Section \ref{Section5} on simulated data.

\begin{SCfigure}
\includegraphics[width=0.6\textwidth]{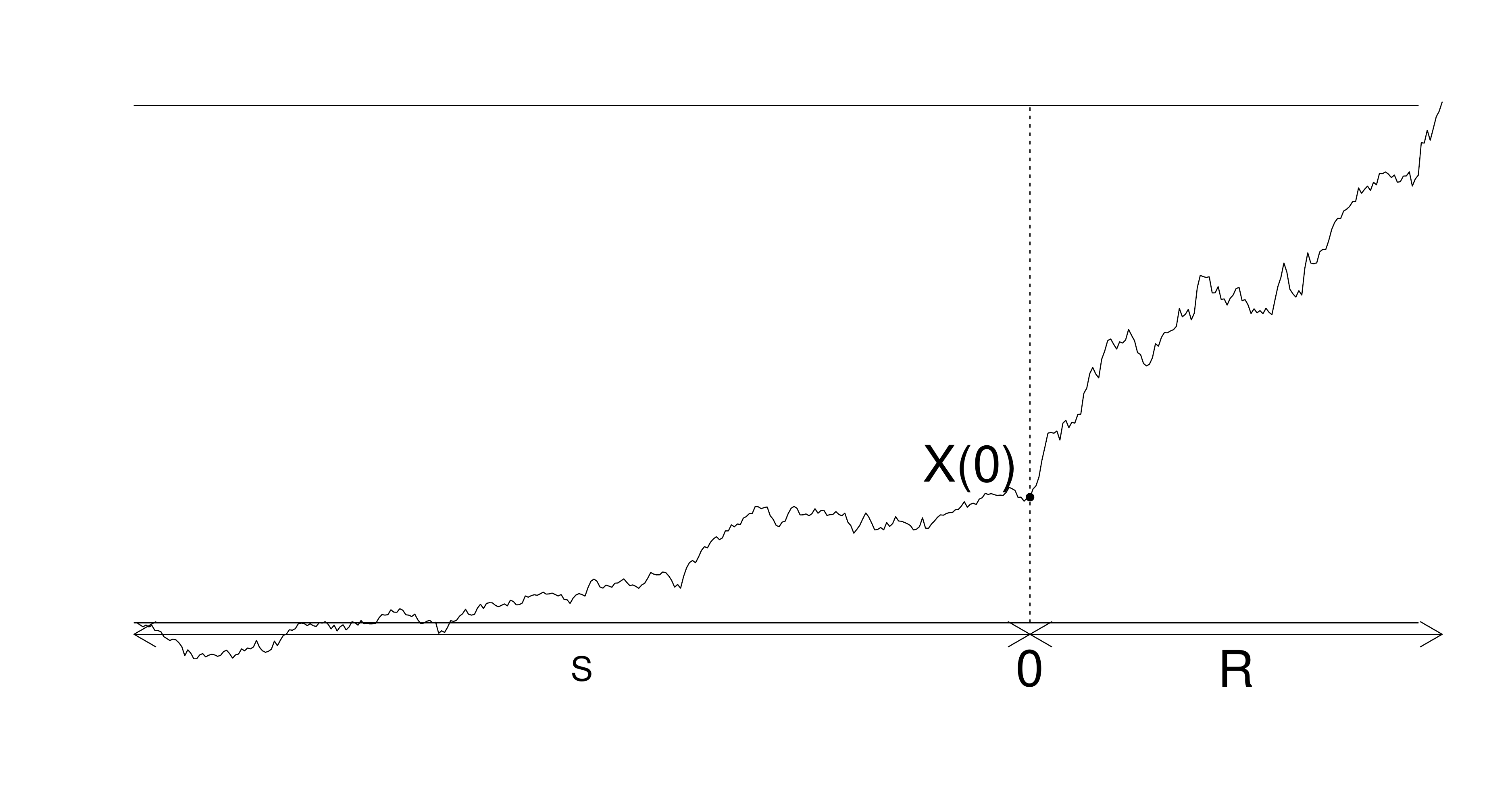}
\caption{Schematic illustration of the single trial. At time $0$, an intervention is initiated, dividing the observed interval into two subintervals: the time $S$ up to the instant of intervention, and the time $R$ between the intervention and the first crossing after it. The random position of the process at time $0$ is denoted by $X(0)$ }
\label{scheme}
\end{SCfigure}

\section{Data}\label{Section2}
The character of experimental data and the description of the involved quantities are illustrated in Fig. \ref{scheme}. At a time independent of when the process started, an intervention is initiated and the time the process has run as well as the time to an event after the intervention are measured. The time of the intervention is set to $0$ by convenience. The intervention divides the observed interval into two subintervals: the time
from the start of the process to the instant of intervention, denoted
by $S$, and the time 
between the intervention and an event after it, denoted by
$R$. Thus, the observed interval has length $S+R$. The experiment is repeated $n$ times. This allows to obtain $n$
independent and identically distributed pairs of intervals
$(S_i,R_i)$, for $i=1,\ldots, n$.  Note that $S_i$ and $R_i$ are not independent.

\section{Model and its properties}\label{Section3}
We describe the dynamics of the system by a diffusion process $X(t)$, starting at some initial value $x_0$.
An event occurs when $X$ exceeds a threshold $B>x_0$ for the first time, which by assumption has not happened before time $0$. The (unobserved) position of the process at the time of the intervention is $X(0)$.   Thus, 
$t$ is running in the interval $[-S,R]$ with $S,R>0$, and we assume $X(t)$ given as
the solution to a stochastic differential equation 
\begin{equation*}\label{model}
\left\{
\begin{array}{l}
dX(t)=\nu \left(X(t),t\right) dt + \sigma\left(X(t),t\right) dW(t),\\
X(-S)=x_0, \qquad X(R)=B,\qquad X(t)<B \textrm{ for } t\in[-S,R),
\end{array}
\right.
\end{equation*}
where $W(t)$ is a standard (driftless) Wiener process. We consider $\nu(X(t),t)=\nu_1\left(X(t)\right)$ and $\sigma(X(t),t)=\sigma_1(X(t))$ for $t<0$, and assume that the intervention causes a change in the parameters of the underlying process to $\nu(X(t),t)=\nu_2(X(t))$, and likewise for $\sigma(X(t),t)$. If there is no intervention, the standard approach is to study the FPT of $X(t)$ through the constant boundary $B$, denoted by $T$. This is the same as the intervention having no effect. Thus, define $T=S+\inf\{ t>0: X(t)\geq B| \nu_1=\nu_2 ,\sigma_1=\sigma_2 \}$. Here $T$ is not observed, but we can still consider its distribution. 

\subsection{Probability densities of $S$, $X(0)$, $R$ and $(S,R)$}  \label{Section4}
It is well known from the theory of point processes that the backward recurrence time $S$ is length biased, and the density is a functional of the distribution of
$T$. In particular, the probability density function (pdf) of $S$ is given
by \citep{CoxLewisBook}, 
\begin{equation}\label{fS}
f_{S}(s)
=\frac{\bar{F}_T(s)}{\mathbb{E}[T]}
\end{equation}
where $\bar{F}_T(s)=1-\mathbb{F}_T(s)=\mathbb{P}(T>s)$ denotes the
survival function, and $\mathbb{E}[T]$ is the mean of $T$. The first two moments of $S$ are given by, \citep{CoxLewisBook},
\begin{equation}\label{moments}
\mathbb{E}[S]=\frac{\mathbb{E}[T^2]}{2\mathbb{E}[T]};\qquad \textrm{Var}[S]=\frac{4\mathbb{E}[T]\mathbb{E}[T^3]-3\mathbb{E}[T^2]^2}{12 \mathbb{E}[T]^2}. \end{equation}
The conditional
density of $X(0)$ given that $B$ has not been crossed up to time $0$ is, 
\citep{aalen},
\begin{equation}\label{fxts}
f_{X(0)}(x|s) =\frac{\frac{\partial}{\partial x} \mathbb{P}(X(0) < x, T>s)}{\mathbb{P}(T>s)}=
\frac{f^a_{X(0)}(x,s)}{\bar{F}_T(s)},
\end{equation}
where $f^a_{X(0)}(x,s)$ denotes the  pdf of the process at time $0$ in presence of a constant absorbing boundary and given that $X(-S)=0$. 
The unconditional density of $X(0)$ is given by 
\begin{equation}\label{fX0}
f_{X(0)}(x) =\int_0^\infty f_{X(0)}(x|s)f_S(s) ds = \frac{1}{\mathbb{E}[T]}\int_0^\infty f^a_{X(0)}(x,s)ds,
\end{equation}
where we used \eqref{fS} and \eqref{fxts}. The variable $R$ coincides with the FPT of $X$ through the boundary $B$, when the process starts in the random position $X(0)<B$ with conditional density $f_{R|X(0)}(r|x)$. The unconditional pdf of $R$ is given by
\begin{equation}\label{fRg}
f_R(r)=\int_{-\infty}^B f_{R|X(0)}(r|x) f_{X(0)}(x) dx.
\end{equation}
The joint pdf of $(S,R)$ is
\begin{equation}\label{count}
f_{(S,R)}(s,r)  
=\frac{1}{\mathbb{E}[T]}\int_{-\infty}^B f_{R|X(0)}(r|x) f^a_{X(0)}(x,s) dx
\end{equation}
since
\begin{eqnarray}
\nonumber F_{(S,R)}(s,r)  
&=&\int_{0}^s \mathbb{P}(R< r| S=u)f_{S}(u)du\\
\nonumber &=&  \int_{0}^s\int_{-\infty}^B \mathbb{P}(R< r|X(0)=x, S=u)f_{X(0)}(x|u)f_{S}(u)dx du\\
\nonumber &=&\int_{0}^s\int_{-\infty}^B \int_{0}^r f_{R|X(0)}(t|x)f_{X(0)}(x|u) f_{S}(u)dt dx du\\
\nonumber &=&\frac{1}{\mathbb{E}[T]}\int_{0}^s\int_{-\infty}^B \int_{0}^r f_{R|X(0)}(t|x) f^a_{X(0)}(x,u)dt dx du,
\end{eqnarray}
where we condition on $X(0)$, then use the Markov property, and finally insert \eqref{fS} and \eqref{fxts}. 

\section{The Wiener process}\label{secWiener}
\begin{figure}
\includegraphics[width=\textwidth]{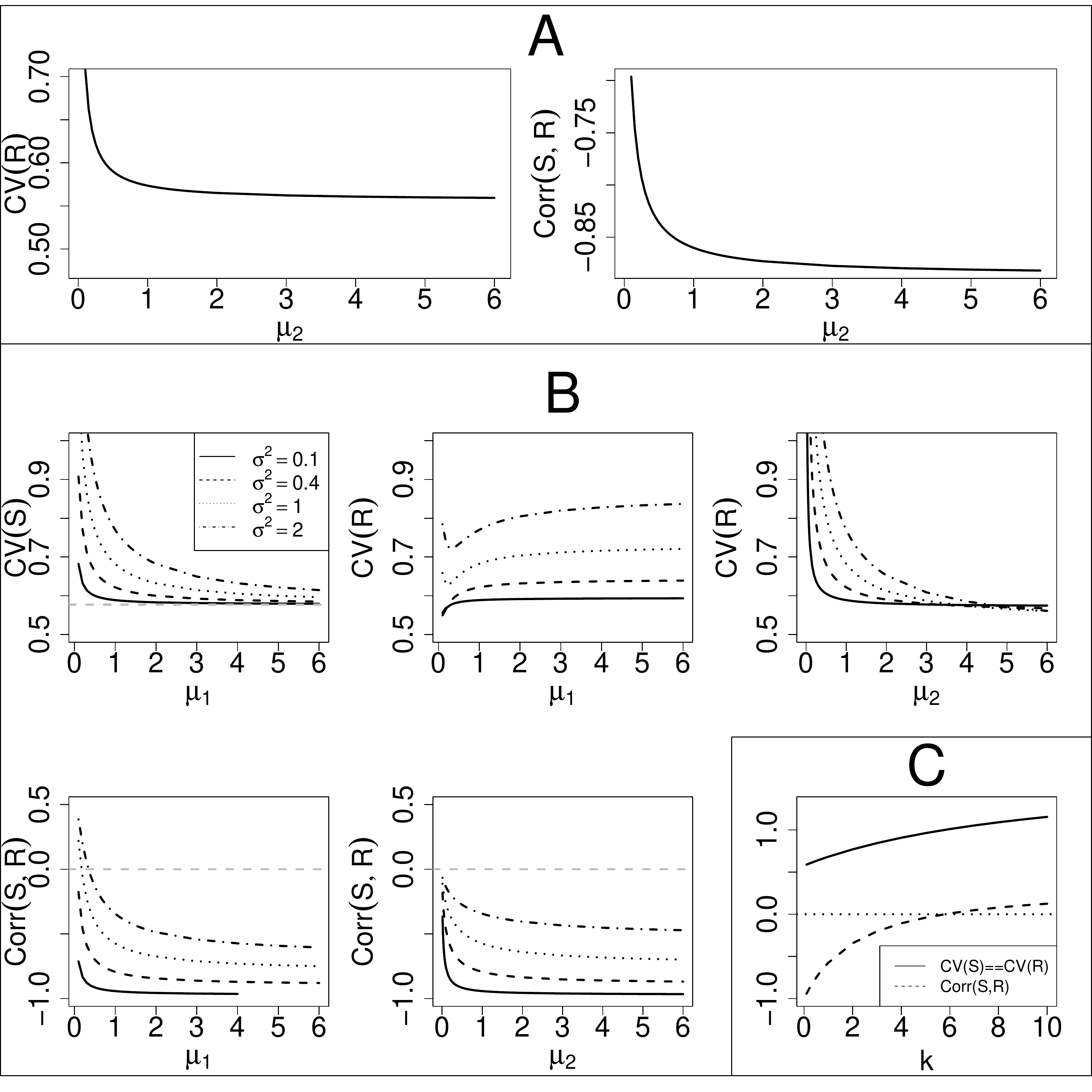}
\caption{Theoretical $\textrm{CVs}$ of $S$ and $R$ and $\textrm{Corr}(S,R)$ as functions of $\mu_1, \mu_2$ and $k$. Panel A) No further assumptions are made. The parameters are $\mu_1=1, \sigma_1^2=0.4, \sigma_2^2=0.1$, yielding an approximate $\textrm{CV}(S)=0.62$. Panel B) Equal variances $\sigma_1^2=\sigma_2^2=0.1, 0.4, 1$ and $2$, the parameters are $\mu_1=1$ for $\mu_2\in [0.1,6]$, yielding an approximate $\textrm{CV}(S)=0.59, 0.62, 0.68$ and $0.77$; $\mu_2=1$ for $\mu_1\in [0.1,6]$. Panel C) The variances are proportional to the drifts, i.e. $\sigma_i^2=k \mu_i, k>0$. The parameters are $\mu_1=1$ and $\mu_2=2$. Note that in this case, $\textrm{CV}(S), \textrm{CV}(R)$ and $\textrm{Corr}(S,R)$ are the same for any value of $\mu_1$ and $\mu_2$, since they do not depend on $\mu_1$ and $\mu_2$ (see Section \ref{Specialcase})}
\label{theoCV}
\end{figure}

Consider a Wiener process $X$ with $\nu_1 (X(t))=\mu_1 > 0$ and $\sigma_1(X(t),t)=\sigma_1>0$ for $t<0$ and assume
that the intervention causes a change in the parameters of the underlying process to $\mu_2, \sigma_2>0$. Because of the space homogeneity, set $x_0=0$ without loss of generality. Since $X$ is a Wiener process with positive drift, $T$ follows an IGD,  $T\sim IG( B/\mu_1,
B^2/\sigma_1^2)$, mean $\mathbb{E}[T] =B/\mu_1$ and variance $\textrm{Var}[T]=B\sigma_1^2/\mu_1^3$ \citep{InverseGaussianBook}. The pdf of $S$ follows from \eqref{fS},
\begin{equation}\label{fSW}
f_S(s)=\frac{\mu_1}{B}\left\{\Phi\left(\frac{B-\mu_1s}{\sqrt{\sigma^2_1 s}}\right)-\exp\left[\frac{2\mu_1 B}{\sigma_1^2}\right]\Phi\left(\frac{-B-\mu_1s}{\sqrt{\sigma_1^2 s}}\right)\right\},
\end{equation}
where $\Phi(\cdot)$ denotes the cumulative distribution function of a standard normal distribution. Inserting the first three moments of $T$ into \eqref{moments}, we get
\begin{equation}\label{momS}
\mathbb{E}[S]=\frac{B\mu_1+\sigma_1^2}{2\mu_1^2};\quad \textrm{Var}[S]=\frac{1}{3}\left(\frac{(B\mu_1+3\sigma_1^2)}{2\mu_1^2}\right)^2; \quad
\textrm{CV}(S)=\frac{B\mu_1+3\sigma_1^2}{\sqrt{3}(B\mu_1+\sigma_1^2)},
\end{equation}
where $\textrm{CV}(S)$ denotes the coefficient of variation of $S$, defined as the ratio between the standard deviation and the mean. 
The pdf of $X(0)$ in presence of a constant absorbing boundary $B$ is \citep{aalen,coxMiller, GiraudoGS,ReviewSac}
\begin{equation}
\label{faW} f^a_{X(0)}(x,s)=\frac{1}{\sqrt{2\pi\sigma^2_1s}}\left\{\exp\left[-\frac{(x-\mu_1 s)^2}{2 \sigma_1^2 s}\right]-\exp\left[\frac{2\mu_1B}{\sigma_1^2}-\frac{(x-2B-\mu_1 s)^2}{2\sigma_1^2 s}\right]\right\},
\end{equation}
for $x\in (-\infty, B)$. Inserting \eqref{faW} into \eqref{fX0}, we get
\begin{equation}\label{fX0W}
f_{X(0)}(x)=\frac{1}{B}\left[\exp\left(\frac{\mu_1(x-|x|)}{\sigma_1^2}\right)-\exp\left(\frac{2\mu_1(x-B)}{\sigma_1^2}\right)\right].
\end{equation}
The mean and variance of $X(0)$ are given by
\begin{equation}\label{meanX0}
\mathbb{E}[X(0)]=\frac{B\mu_1-\sigma_1^2}{2\mu_1}, \qquad \textrm{Var}[X(0)]=\frac{B^2\mu_1^2+3\sigma_1^4}{12\mu_1^2}.
\end{equation}
The distribution of $R$ conditioned on $X(0)=x$ is  $R|X(0)\sim IG\left((B-x)/\mu_2,(B-x)^2/\sigma_2^2\right)$. Plugging this and \eqref{fX0W} into \eqref{fRg}, we obtain
\begin{eqnarray*}
f_R(r)&=&\frac{\mu_2}{B}\left[\Phi\left(\frac{B-\mu_2r}{\sigma_2\sqrt{r}}\right)-\Phi\left(-\frac{\mu_2 \sqrt{r}}{\sigma_2}\right)\right]
+\frac{\mu_2\sigma_1^2-2\mu_1\sigma_2^2}{B\sigma_1^2}\exp\left(\frac{2\mu_1r(\mu_1\sigma_2^2-\mu_2\sigma_1^2)}{\sigma_1^4}\right)\\
&\times&\left[\exp\left(\frac{2\mu_1B}{\sigma_1^2}\right)\Phi\left(-\frac{B\sigma_1^2+2r\mu_1\sigma_2^2-\mu_2 r\sigma_1^2}{\sigma_1^2\sigma_2\sqrt{r}}\right)-\Phi\left(-\frac{2\mu_1r\sigma_2^2-\mu_2r\sigma_1^2}{\sigma_1^2\sigma_2\sqrt{r}}\right)\right].
\end{eqnarray*}
Finally, using \eqref{faW} and $f_{R|X(0)}$ in \eqref{count}, we get 
\begin{eqnarray}
\nonumber && f_{(S,R)}(s,r)= 
\frac{\mu_1}{B\sqrt{2\pi[\sigma_1^2s +\sigma_2^2 r]^3}}
\exp\left\{-\frac{(B-\mu_1s-\mu_2r)^2}{2(\sigma_1^2s + \sigma_2^2r)}\right\}\\
\nonumber&\times& \left\{[(B-\mu_1s)\sigma_2^2 +s\mu_2\sigma_1^2]\Phi\left(\sqrt{r}\frac{(B-\mu_1s)\sigma_2^2+s\mu_2\sigma_1^2}{\sigma_1\sigma_2\sqrt{s(\sigma_1^2s+\sigma_2^2r)}}\right)\right.\\
\nonumber &-&\left.\exp\left\{\frac{2rB(\mu_1\sigma_2^2-\mu_2\sigma_1^2)}{\sigma_1^2 (\sigma_1^2 s+\sigma_2^2 r)}\right\}[(-B-\mu_1s)\sigma_2^2+\mu_2\sigma_1^2s]\Phi\left( \frac{(-B-\mu_1s)\sigma_2^2+\mu_2\sigma_1^2 s}{\sigma_1\sigma_2\sqrt{s(\sigma_1^2s+\sigma_2^2r)}}\sqrt{r}\right)\right\}.\\
[-.5ex]\label{fTXY}
\end{eqnarray}
No closed expressions for $\textrm{CV}(R)$, covariance and correlation of $S$ and $R$ are available, except for $\sigma_i^2=k \mu_i, k>0$, as described below. In Fig. \ref{theoCV} we illustrate $\textrm{CV}(S)$ given by \eqref{momS} and numerically approximate $\textrm{CV}(R), \textrm{Cov}(S,R)$ and $\textrm{Corr}(S,R)$ for those parameter values used in Section \ref{Section5}. Note that when $\mu_2\to \infty$, the expected time for an event after the intervention goes to zero; $\mathbb{E}[R]\to 0$. Also, $\textrm{Var}[R]\to 0$, whereas $\textrm{CV}(R)$ does not, as shown in Fig. \ref{theoCV}.

\subsection{Special case:  squared diffusion coefficients proportional to the drifts}\label{Specialcase}
Up to now, we made no assumptions on the relation between changes in the drift and changes in the variance of the Wiener process. However, in many applications larger values of a variable are followed by a larger variance. This is formalized, for example, by the well known psychophysical Weber's law, claiming that the standard
deviation of the signal is proportional to its strength \citep{Laming}. Applying this law to the IGD by relating mean and standard deviation, given prior to eq. \eqref{fSW}, we obtain that $\sigma^2$ is proportional to $\mu$. An analogous result can be derived from the diffusion approximation procedure \citep{LanskySac01}. We therefore assume the squared diffusion coefficients proportional to the drift coefficients, i.e. $\sigma^2_i=k \mu_i$, for $k>0, i=1,2,$. The above expressions simplify to  
\begin{eqnarray}
\label{specialS}\mathbb{E}[S]&=&\frac{B+k}{2\mu_1},\quad \textrm{Var}[S]=\frac{(B+3k)^2}{12\mu_1^2}, \quad \textrm{CV}(S)=\frac{B+3k}{\sqrt{3}(B+k)},\\
\nonumber\mathbb{E}[X(0)]&=&\frac{B-k}{2}, \quad \textrm{Var}[X(0)]=\frac{B^2+3k^2}{12},\\
\nonumber f_R(r)&=&\frac{\mu_2}{B}\left\{\Phi\left(\frac{B-\mu_2r}{\sqrt{k\mu_2 r}}\right)-\exp\left(\frac{2 B}{k}\right)\Phi\left(\frac{-B-\mu_2r}{\sqrt{k \mu_2 r}}\right)\right\}=\frac{\bar {F}_{T^*}(r)}{\mathbb{E}[T^*]},
\end{eqnarray}
where  $T^*$ denotes the FPT through $B$ of the Wiener process starting in $0$ with drift $\mu_2$ and diffusion coefficient $\sqrt{k\mu_2}$.  Note that $R$ is distributed as the forward recurrence time of $T^*$, as well as $S$ is distributed as the backward recurrence time of $T$. Thus
\begin{equation}
\label{specialR} \mathbb{E}[R]=\frac{B+k}{2\mu_2},\qquad \textrm{Var}[R]=\frac{(B+3k)^2}{12\mu_2^2},\qquad   \textrm{CV}(R)=\frac{B+3k}{\sqrt{3}(B+k)}.
\end{equation}
Interestingly, $\textrm{CV}(S)=\textrm{CV}(R)$ and they only depend on $k$ and not on the specific values of the coefficients. The joint pdf of $S$ and $R$ is
\begin{eqnarray}
\nonumber f_{(S,R)}(s,r)&=&\frac{\mu_1\mu_2}{\sqrt{2\pi k (\mu_1s+\mu_2r)^3}}\exp\left(-\frac{(B-\mu_1s-\mu_2r)^2}{2k(\mu_1s+\mu_2r)}\right)\\
\label{fSRb} &=&\frac{\mu_1\mu_2}{B}f_{IG(B,B^2/k)}(\mu_1s+\mu_2r),
\end{eqnarray}
and the covariance and correlation of $S$ and $R$ are 
\begin{eqnarray}
\label{CovCorr}\textrm{Cov}(S,R)&=&\mathbb{E}[SR]-\mathbb{E}[S]\mathbb{E}[R]=\frac{3k^2-B^2}{12\mu_1\mu_2},\\
\label{Corr}\textrm{Corr}(S,R)&=&\frac{\textrm{Cov}(S,R)}{\sqrt{\textrm{Var}[S]\textrm{Var}[R]}}=\frac{3k^2-B^2}{(B+3k)^2},
\end{eqnarray}
see Appendix \ref{Appendix1}. Note that the correlation can be positive, null or negative, depending on whether $0<k<B/\sqrt{3}, k=B/\sqrt{3}$ or $k>B/\sqrt{3}$, respectively. Moreover, $\textrm{Corr}(S,R)\to-1$ as $k\to 0$, i.e. $\sigma_i^2\to 0$, while $\textrm{CV}(S)=\textrm{CV}(R)\to \sqrt 3$ and $\textrm{Corr}(S,R)\to 1/3$ as $k\to \infty$, i.e. $\sigma_i^2\to \infty, i=1,2$.

\section{Parameter estimation}\label{Section5}
The aim of this paper is the estimation of the parameters of $X$ from a sample $\{(s_i,r_i)\}_{i=1}^n $ of $n$ independent observations of $(S,R)$, and testing if the intervention has an effect by the hypothesis $H_0: \mu_1=\mu_2$. Three scenarios are considered: no information about the parameters is available; we assume equal variances $\sigma_1^2=\sigma_2^2=\sigma^2$; or we assume $\sigma^2_i=k\mu_i$, as in Section \ref{Specialcase}. That is, we want to estimate either $\phi=(\mu_1,\sigma_1^2,\mu_2,\sigma_2^2), \phi=(\mu_1,\mu_2,\sigma^2)$ or $\phi=(\mu_1,\mu_2,k)$. Since the $(s_i, r_i)$'s, $i=1,\ldots, n$ are independent and identically distributed, the log-likelihood is $l_{(s,r)}(\phi)=\sum_{i=1}^n \log f_{(S,R)}(s_i,r_i)$.
The maximum likelihood estimator $\hat\phi$ is found by maximizing $l_{(s,r)}$ numerically (see Appendix \ref{Appendix2}). An approximate 95\% confidence interval (CI) for $\phi_i$ is given by $\hat\phi_i \pm 1.96\ \textrm{SE}(\hat\phi_i)$, where $\textrm{SE}$ is the asymptotic standard error given by $\textrm{SE}(\hat\phi_i)=\sqrt{I_{ii}(\hat\phi)^{-1}/n}$,  where $I(\phi)$ is the Fisher information matrix
 \citep{Cramer}, which we  approximate  numerically (see Appendix \ref{Appendix2}). To test the hypothesis $H_0:\mu_1=\mu_2$ we perform a likelihood ratio test at a $5\%$ significance level, evaluating it in a chi-squared distribution with one degree of freedom. We reject $H_0$ if $-2\log [ L_0(\hat\phi_0)/L_\textrm{full}(\hat\phi)]> 3.84$, where $L_0$ and $L_\textrm{full}$ denote the likelihood functions of the null and full (alternative) model evaluated in the estimated parameters $\hat\phi_0=(\hat\mu,\hat\sigma^2)$ and $\hat\phi=(\hat\mu_1,\hat\mu_2,\hat\sigma^2)$ under the hypotheses $\mu=\mu_1=\mu_2$ and $\mu_1\neq \mu_2$, respectively. This test can be applied in all the considered scenarios, but for simplicity we only report results for the case of equal variances. Results for the other cases are similar.  We assume both the parametric form of the underlying process and the relations between parameters, if any, to be known. It can be discussed if these assumptions are realistic. Equality of diffusion coefficients, or the assumption of variance proportional to the mean, can be checked by likelihood ratio test.

\subsection{Monte Carlo simulation study}
For the simulations, parameter values are chosen such that the mean of $T$ in the case of no intervention is five times its standard deviation. This is obtained by setting $B=10, \mu_1=1, \sigma_1^2=0.4$, yielding $\mathbb{E}[T]=10$, and $\textrm{Var}[T]=4$. Then $\mu_1$ and $\sigma_1^2$ are varied to investigate different regimes of the model. Also the effect of the intervention is varied through the parameters $\mu_2$ and $\sigma_2^2$. Samples of size $n=100$ are simulated, and for each set of parameter values, we repeat simulation of data set and estimation 1000 times, obtaining 1000 statistically independent trials. 

We calculated coverage probabilities (CPs), defined as the probability that the CI covers the true value, to evaluate the performance of the CIs. The CP should be close to $1-\alpha$, where $\alpha=0.05$ is the significance level, and the CI should be narrow for a reliable estimator.

The computing environment \textbf{R} has been used to carry out both the simulations of $(s_i,r_i)$ and the parameter estimation. A description of the simulation procedure is reported in Appendix \ref{Appendix3}.

\begin{table}
\centering
  \resizebox{\linewidth}{!}{
\begin{tabular}{|c||c|c|c|c||c|c|c|c|}
\hline
&Average & Empirical & Asymptotic& & Average & Empirical & Asymptotic & \\ 
CV(R) & $\textrm{of } \hat\mu_1$& $ \textrm{SE}(\hat\mu_1)$& $ \textrm{SE}(\hat\mu_1)$&$\textrm{CP}(\hat\mu_1)$&$ \textrm{of } \hat\sigma_1^2$ &  $ \textrm{SE}(\hat\sigma^2_1)$& $ \textrm{SE}(\hat\sigma^2_1)$&$\textrm{CP}(\hat\sigma^2_1)$\\ \hline
0.60&0.9998& 0.0405& 0.0397& 94.7& 0.39962& 0.1079 &0.1027& 91.6\\
0.65& 1.0020& 0.0438& 0.0428&93.7 &0.4016 &0.1213 &0.1154 &91.3\\
0.70& 1.0023 &0.0468 &0.0441 &94.5 &0.3983 &0.1315& 0.1198 &91.8\\
0.75& 1.0020 &0.0458 &0.0449 &94.9 &0.3989 &0.1388 &0.1251& 91.4\\
\hline
\multicolumn{7}{} \hline \\ 
\hline
&Average & Empirical & Asymptotic& & Average & Empirical & Asymptotic & \\ 
CV(R) & $\textrm{of } \hat\mu_2$& $ \textrm{SE}(\hat\mu_2)$& $ \textrm{SE}(\hat\mu_2)$&$\textrm{CP}(\hat\mu_2)$&$ \textrm{of } \hat\sigma_2^2$ &  $ \textrm{SE}(\hat\sigma^2_2)$& $ \textrm{SE}(\hat\sigma^2_2)$&$\textrm{CP}(\hat\sigma^2_2)$\\ \hline
0.60& 0.1003& 0.0032& 0.0032& 94.8& 0.0256& 0.0083& 0.0080& 92.7\\
0.65& 0.1001 &0.0044 &0.0043 &93.7 &0.0578& 0.0154 &0.0145 &91.9\\
0.70& 0.1000 &0.0053 &0.0051 &93.7& 0.0926& 0.0221& 0.0212& 92.1\\
0.75& 0.1001 &0.0058 &0.0058 &95.5& 0.1290& 0.0288& 0.0278& 92.9\\ \hline
\end{tabular}}
\caption{Averages, empirical and  asymptotic SEs and
  CPs  in percentage over 1000 estimates of
  $\phi=(\mu_1,\sigma_1^2,\mu_2,\sigma_2^2)$ for $n=100$, when  $\mu_1=1,
  \sigma_1^2=0.4, \mu_2=0.1$, and $\sigma_2^2 = 0.026, 0.059, 0.094$,
  or $0.131$,
  yielding an approximate $\textrm{CV}(R)=0.60, 0.65, 0.70$ or
  $0.75$, respectively. In all cases, $\textrm{CV}(S)=0.62$. } 
\label{table1}
\end{table}

\begin{figure}
\includegraphics[width=\textwidth]{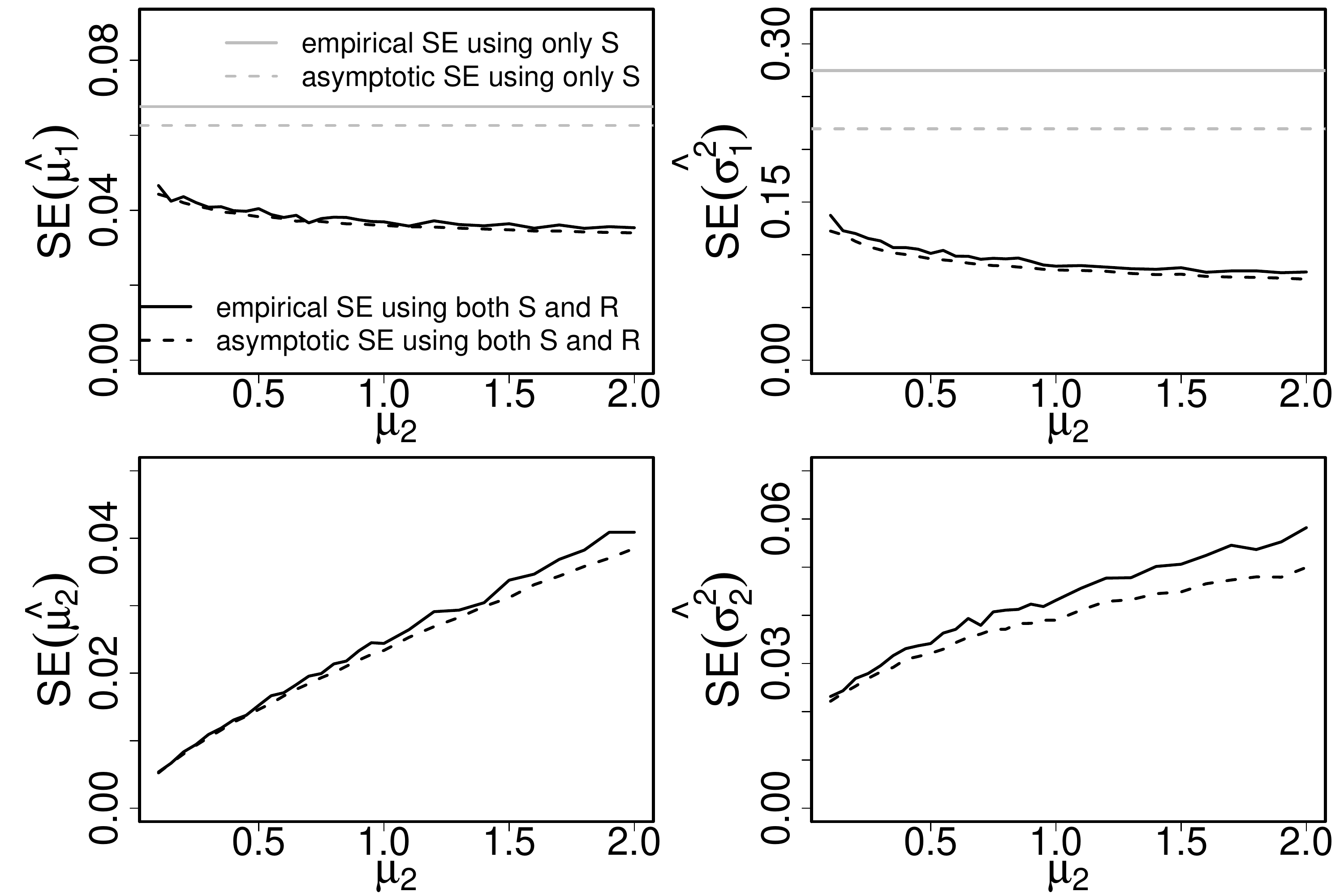}
\caption{Empirical and asymptotic SEs over 1000 estimates of $(\mu_1,\sigma_1^2,\mu_2,\sigma_2^2)$ for $n=100$ as a function of $\mu_2$ when no assumptions on the parameters are made. The parameters are $\mu_1=1, \sigma_1^2=0.4, \sigma_2^2=0.1$, yielding an approximated $\textrm{CV}(S)=0.62$. Full lines: empirical SEs. Dashed lines: asymptotic SEs. Colors correspond to the SEs of the estimators obtained by either maximizing $l_{(S,R)}$ (black lines), or maximizing $\log f_S$ (gray lines), respectively }
\label{caseaoggi}
\end{figure}

\subsection*{No further assumptions on parameters}
We choose $\mu_1=1, \sigma_1^2=0.4$ and thus $\textrm{CV}(S)=0.62$. First we fix $\mu_2$ and vary $\sigma_2^2$, then we fix $\sigma_2^2$ and let $\mu_2$ vary. In the first case, we fix $\mu_2=0.1$, implying that the  intervention slows down the process, since $\mu_2< \mu_1$. To obtain $\textrm{CV}(R)= 0.6, 0.65, 0.7$ or $0.75$, we set $\sigma_2^2=  0.026, 0.059, 0.094$, or $0.131$. 

Averages and empirical SEs of the estimates, as well as medians of the asymptotic SEs and the CPs of the  CIs  are reported in Table \ref{table1}. All estimators appear unbiased and with acceptable SEs. 
The empirical and asymptotic SEs are approximately equal, suggesting that $n=100$ is sufficient for asymptotics to be valid. Not surprisingly, the performance improves when the CV of $R$ decreases.  This holds also for $\hat\mu_1$ and $\hat\sigma_1^2$, highlighting the dependence between $S$ and $R$: a
large variability after the intervention deteriorates estimation of parameters governing the process before the intervention. All CPs are close to the desired $95\%$. The CPs of $\mu_1$ and $\mu_2$ are higher than those of $\sigma_1^2$ and $\sigma_2^2$. This  phenomenon disappears for larger $n$, when all CPs are around $95\%$ (results not shown). 

In the second case, we let $\mu_2$ vary in the interval $[0.1,10]$, and fix $\sigma_2^2=0.1$.  Here the response to the intervention either slows down or accelerates the process, depending on whether $\mu_2< \mu_1$ or $\mu_1< \mu_2$, respectively.

A relevant question is how much, if at all, the estimators of $\mu_1$ and $\sigma_1^2$ improve by considering the more complicated likelihood based on  eq. \eqref{fTXY} compared to the simple likelihood based on eq. \eqref{fSW}, where information from $R$ is ignored. All estimators appear unbiased (figures not shown). The estimates of $\mu_1$ and $\sigma_1^2$ obtained from observations of $(S,R)$ outperform those obtained only from observations of $S$, as can be seen comparing both their empirical and asymptotic SEs in Fig. \ref{caseaoggi}. When $\mu_2$ increases, the performance of $\hat\mu_1$ and $\hat\sigma^2_1$ improve and that of $\hat\mu_2$ and $\hat\sigma_2^2$ get worse even if $\textrm{CV}$ of $R$ decrease.  Moreover, the empirical and asymptotic SEs for $\hat\mu_2$ and $\hat\sigma_2^2$ are quite different for large $\mu_2$, e.g. $\mu_2=10, \mu_2=500$, meaning that $n=100$ is not sufficient for asymptotics to be valid. In the other cases the empirical and asymptotic SEs are approximately equal, and thus in the following we only report the asymptotic values.

\begin{figure}
\centering \includegraphics[width=\textwidth]{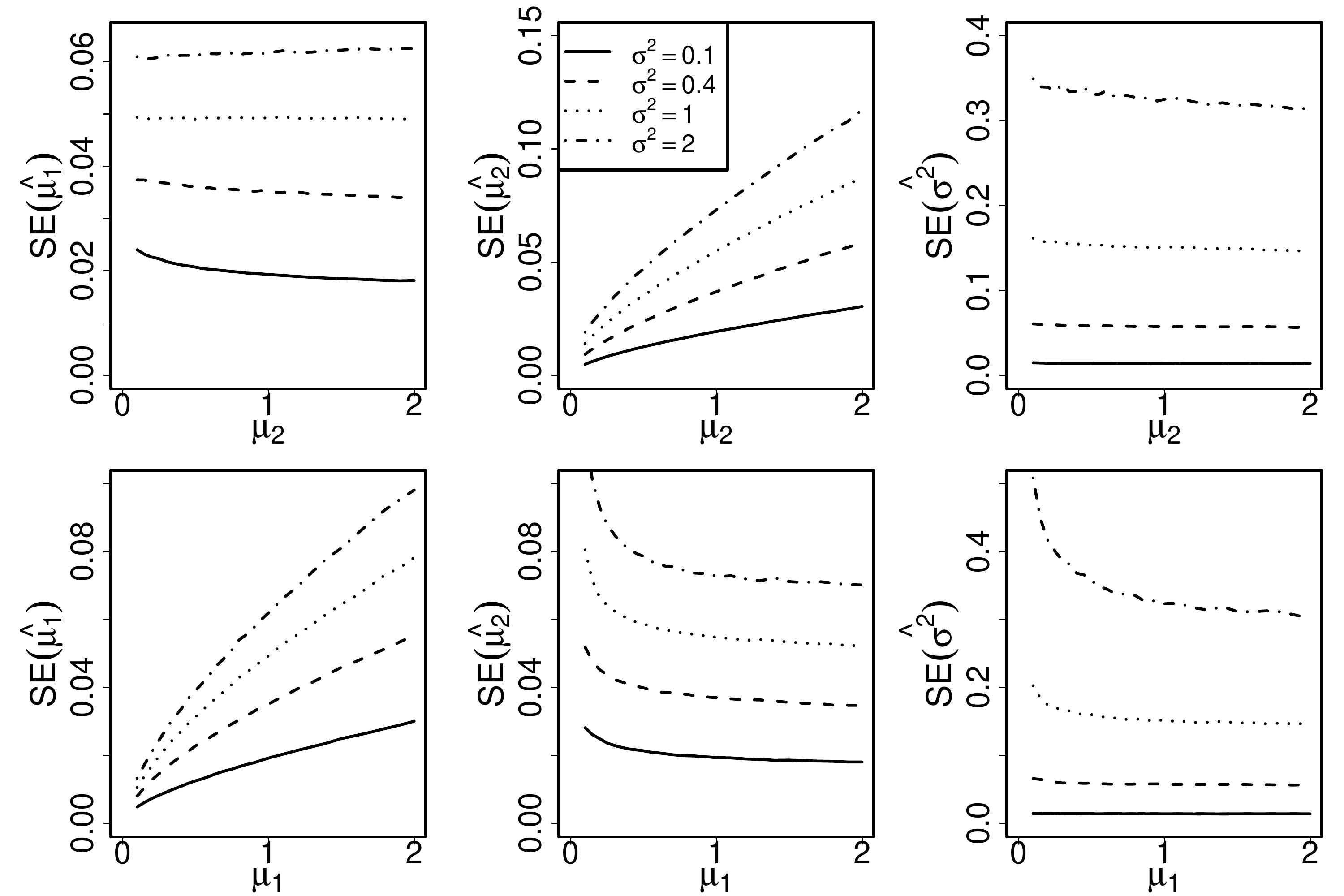}
\caption{Asymptotic SEs  over 1000 estimates of $(\mu_1,\mu_2,\sigma^2)$ for $n=100$ as a function of $\mu_2$ (upper panels) and of $\mu_1$ (lower panels) for equal variances, $\sigma_1^2=\sigma_2^2=\sigma^2.$  In both cases, $\sigma^2=0.1$ (full lines), $0.4$ (dashed lines), $1$ (dotted lines) and $2$ (dotted-dashed lines). In the upper panel, $\mu_1=1$ (upper panels) yielding an approximate $\textrm{CV}(S)=0.59, 0.62, 0.68$ and $0.77$, respectively, and in the lower panel $\mu_2=1$  }\label{Figba}
\end{figure}

\begin{figure}
\centering \includegraphics[width=\textwidth]{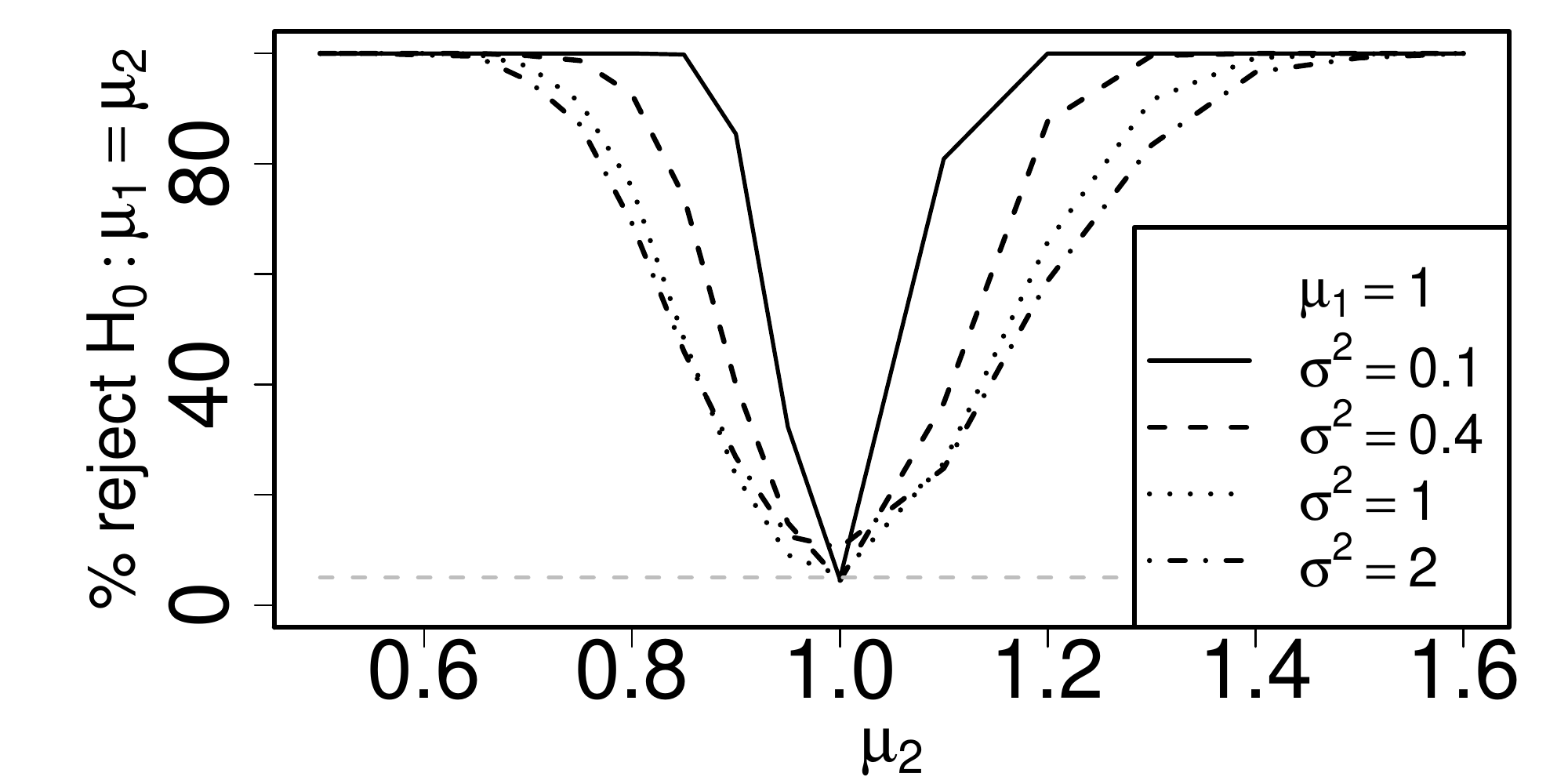}
\caption{Percentage of rejections, using the likelihood ratio test at $5\%$ significance level of the null hypothesis $H_0: \mu_1=\mu_2$ as a function of $\mu_2$ for equal variances, $\sigma^2=\sigma_1^2=\sigma_2^2$. The parameters are $\mu_1=1, \sigma^2=0.1$ (full line), 0.4 (dashed line), 1 (dotted line) and 2 (dashed-dotted line) }
\label{test}
\end{figure}

\subsection*{Equal variances}
When $\sigma_1^2 = \sigma_2^2=\sigma^2$, we put $\sigma^2=0.1, 0.4, 1$ and $2$, respectively, with either $\mu_1=1$ and $\mu_2 \in [0.1, 10]$ or $\mu_2=1$ and $\mu_1\in [0.1,10]$. The variability of the estimators  for different values of $\mu_1$ and $\mu_2$  is reported in Fig. \ref{Figba}, where the SEs of the estimators are plotted against $\mu_2$. The estimators appear unbiased (results not shown). All of them improve when $\sigma^2$ decreases, since that reduces the variability of both $S$ and $R$. The performance of $\hat\mu_i$ improves while that of $\hat\mu_j$ gets worse when $\mu_j$ increases, for $i,j=1,2$ and $i\neq j$. Interestingly, the performance of $\hat\sigma^2$ seems to be constant with respect to $\mu$, unless $\sigma^2$ is large. 

A likelihood ratio test is performed for testing the hypothesis $H_0:\mu_1=\mu_2$ at a $5\%$ significance level and the percentage of rejections of $H_0$ as a function of $\mu_2$ is reported in Fig. \ref{test}. If $\mu_1=\mu_2$, we want the percentage to be around $5\%$, while if $\mu_1\neq \mu_2$, the percentage represents the power of the test, i.e. the probability of correctly rejecting the null hypothesis, and we want it as high as possible. When $\mu_1=\mu_2=1$, this percentage is around $5\%$ for $\sigma^2=0.1, 0.4$ and $1$, suggesting that $n=100$ is sufficient for asymptotics to be valid. When $\sigma^2=2$, the percentage is $10.4$ and then a larger $n$ should be considered. Not surprisingly, the power of the test decreases when $\sigma^2$ increases, but it is worthwhile noting that it is larger than $50\%$ when $|\mu_1-\mu_2|>0.2$ and around $100\%$ if $|\mu_1-\mu_2 |\geq  0.4$, indicating a satisfactory performance of the test.


\subsection*{Variance proportional to the mean}
\begin{figure}
\centering \includegraphics[width=\textwidth]{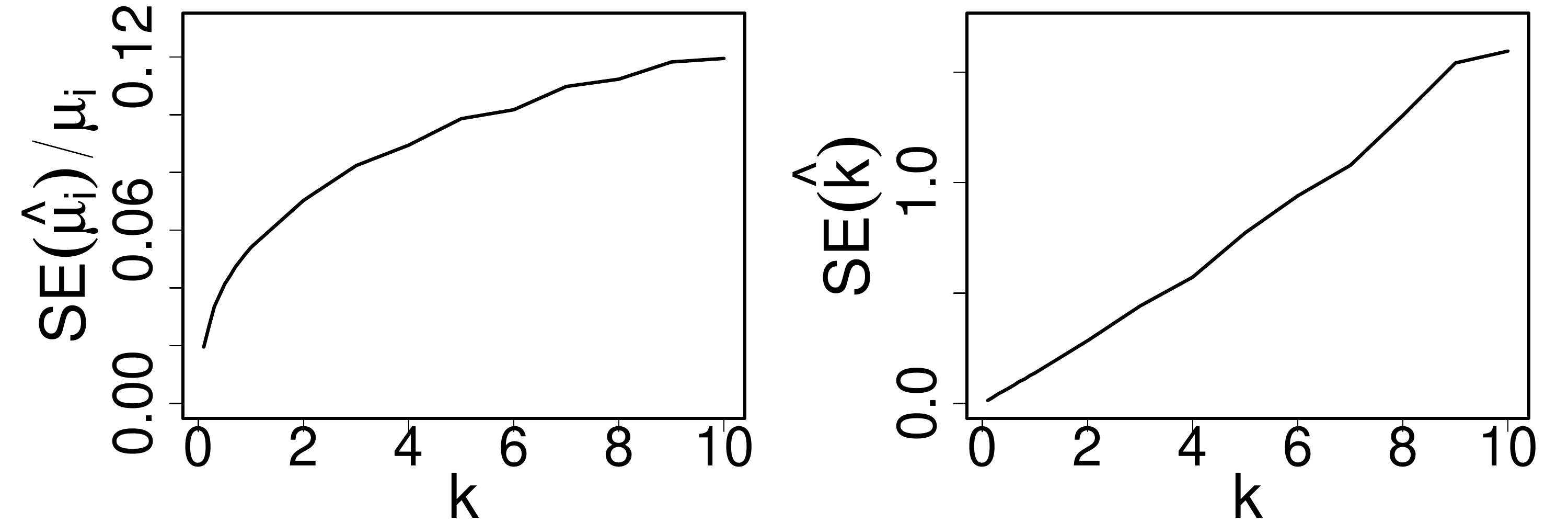}
\caption{Asymptotic SEs over 1000 estimates of $\mu_i$ and $k$ for $n=100$ rescaled by $\mu_i$ as a function of $k$ when the variance is proportional to the mean, $\sigma_i^2=k \mu_i, i=1,2$. The parameters are $\mu_1=1$ and $\mu_2=2$. The results for $\textrm{SE}(\hat\mu_1)/\mu_1$ and $\textrm{SE}(\hat\mu_2)/\mu_2$ are almost indistinguishable. The same results hold for other combinations of $(\mu_1,\mu_2)$ and are therefore not reported } 
\label{Figc}
\end{figure}

Now assume $\sigma_i^2 = k \mu_i$, for $k>0$. The parameter values are $k \in [0.1, 10]$ and $\mu_1, \mu_2 \in \{0.1, 1, 2\}$. The performance of the estimators is reported in Fig. \ref{Figc}, where $\textrm{SE}(\hat\mu_i)/\mu_i$ and $\textrm{SE}(\hat k)$ are plotted against $k$. Also in this case, estimators appear unbiased (results not shown).  As expected from  the theoretical results in Section \ref{Specialcase}, the performance of  $\hat\mu_1$ and $\hat\mu_2$ appears similar, and it does not depend on $\mu_2$ and $\mu_1$, respectively. Interestingly, the asymptotic SE of $\hat k$ depends neither on $\mu_1$ nor on $\mu_2$, but only on $k$. This may be due to the fact that neither the $\textrm{CVs}$ of $S$ and $R$ nor their correlation depend on $\mu_1$ and $\mu_2$, see eqs. \eqref{specialS}, \eqref{specialR} and \eqref{Corr}.

\section{Conclusion}\label{Section6}
When any intervention is applied, the most natural question arising is about its effect. Here, the effect is reflected in the change of the time to an observable event. However, there is no apparent information available about what such a time would be if no action is taken. In this paper we solve the problem by comparing time to the intervention and the time to the final event. The parameters of the underlying system are both identified and statistically compared to judge the presence of an effect. The method represents a potential tool in all the experimental situations where direct measurements are not available, but only the qualitative changes are observable.

\section*{Acknowledgements} 
 S.D. was supported by the Danish Council for Independent Research\,$|$\,Natural Sciences. P.L. supported by grant
No. AV0Z50110509. The work is part of the Dynamical Systems Interdisciplinary Network, University of Copenhagen. 
\vspace{1cm}

{\bf \noindent Appendix}
\appendix
\section{Covariance and Correlation of $S$ and $R$ when $\sigma_i^2=k\mu_i$}\label{Appendix1}
Let $P\sim IG(B,B^2/k)$, and thus $\mathbb{E}[P]=B$. Then, using \eqref{fSRb}, we have
\begin{eqnarray}
\nonumber\mathbb{E}[SR]&=&\int_0^\infty \int_0^\infty sr f_{(S,R)} dr ds = \int_0^\infty \frac{s\mu_1 }{B} \int_0^\infty \mu_2 r f_{P}(\mu_1s+\mu_2r)dr ds\\
\nonumber&=&\int_0^\infty \frac{s\mu_1}{B}\int_{\mu_1s}^\infty \frac{1}{\mu_2} (t-\mu_1s) f_P(t) dt ds\\
\nonumber&=&\frac{1}{\mu_1\mu_2B}\int_0^\infty u\int_u^\infty (t-u)f_P(t)dt du\\
\label{calc1}&=&\frac{1}{\mu_1\mu_2B}\int_0^\infty u\int_u^\infty tf_P(t)dt du-\frac{1}{\mu_1\mu_2B}\int_0^\infty u^2\bar{F}_p(u) du.
\end{eqnarray}
Calculating the integral in \emph{dt} by parts, we get
\begin{equation}\label{calc2}
\int_u^\infty t f_P(t) dt=[-t\bar F_P(t)]|_u^\infty+\int_u^\infty \bar F_p(t) dt=u\bar F_P(u)+\int_u^\infty \bar F_p(t) dt,
\end{equation}
where $-t\bar F_P(t)\to 0$ when $t\to\infty$ because $\bar F(t)=o(t^{-1})$ as $t\to \infty$. Define now a variable $Q$ by
\[
f_Q(t)=\frac{\bar F_P(t)}{\mathbb{E}[P]}=\frac{\bar F_P(t)}{B}.
\]
Then, inserting \eqref{calc2} into \eqref{calc1} and simplifying the resulting expression, we obtain
\begin{equation}\label{calc3}
\mathbb{E}[SR]=\frac{1}{\mu_1\mu_2 B}\int_0^\infty u \int_u^\infty \bar F_P(t) dt du=\frac{1}{\mu_1\mu_2}\int_0^\infty u \int_u^\infty f_Q(t)dt du=
\frac{1}{\mu_1\mu_2}\int_0^\infty u \bar F_{Q}(u)du.
\end{equation}
Similarly, let $Z$ be a variable defined by 
\[
f_Z(u)=\frac{\bar F_Q(u)}{\mathbb{E}[Q]}.
\]
Then \eqref{calc3} becomes
\begin{equation}\label{calc4}
\mathbb{E}[SR]=\frac{\mathbb{E}[Q]}{\mu_1\mu_2}\int_0^\infty u \frac{\bar F_Q(u)}{\mathbb{E}[Q]}du=
\frac{\mathbb{E}[Q]}{\mu_1\mu_2} \mathbb{E}[Z],
\end{equation}
where 
\[
\mathbb{E}[Z]=\frac{1}{2}\mathbb{E}[Q]+\frac{1}{2}\frac{\textrm{Var}[Q]}{\mathbb{E}[Q]},
\]
see eqs. \eqref{fS} and \eqref{moments}. 
Mimicking the calculations done for $S$ in \eqref{specialS}, we obtain $\mathbb{E}[Q]=(B+k)/2, \textrm{Var}[Q]=(B+3k)^2/12$. Plugging them into $\mathbb{E}[Z]$ first and then \eqref{calc4}, and simplifying the resulting expression, we get
\[
\mathbb{E}[SR]=\frac{B^2+3Bk+3k^2}{6 \mu_1\mu_2}.
\]
Finally, \eqref{CovCorr} follows  using \eqref{specialS} and \eqref{specialR}.

\section{Simulation in \textbf{R} }\label{Appendix3}
To simulate $(s_i,r_i), i=1,\ldots, n$  we proceed as follows. We simulate $s_i$ by applying the inverse transforming sampling to the cumulative distribution function of $S$, which is obtained by numerically integrating \eqref{fS} using the function {\tt integrate} in \textbf{R}. We obtain $s_i$ by simulating $u_i$ from a uniform distribution on $[0,1]$, and solving $F_{S}(s_i)-u_i=0$ with respect to $s_i$ by means of the function {\tt uniroot} in \textbf{R}. 
To obtain an observation $r_i$ from $R$ we first simulate $x$, i.e. the position $X(0)$ of the process at the time of intervention. We use the inverse transforming sampling to the distribution of $X(0)$, obtained by integrating \eqref{fxts} with respect to $x$, i.e. $F_{X(0)}(x|s)=F^a_{X(0)}(x,s)/\mathbb{P}(T>s)$.
Because $X$ is a Wiener process, $F^a_{X(0)}(x,s)$ is given by \eqref{faW},
\[
F^a(x,s)= \Phi\left(\frac{x-\mu_1s}{\sqrt{\sigma^2_1 s}}\right)-\exp\left[\frac{2\mu_1B}{\sigma_1^2}\right]\Phi\left(\frac{x-2B-\mu_1s}{\sqrt{\sigma^2_1 s}}\right).
\]
Using $x$, an observation $r_i$ from $R$ is drawn from $IG((B-x)/\mu_2,(B-x)^2/\sigma_2^2)$. 

\section{Estimation of $\phi$ and $I(\phi)$ in \textbf{R}}\label{Appendix2}
Since all parameter values need to be positive, maximizing the log-likelihood  is a constrained optimization problem.  However, the estimated parameters are always positive when estimating $\phi$ simply by minimizing $-l_{(s,r)}$ by means of the function {\tt optim}. 

Since $l_{(s,r)}$ is a complicated function of $\phi$, it can frequently happen that it has several local maxima. To find the global maximum, sensible starting values are paramount. The starting value $\phi_0$ for the iterations is chosen by the following strategy:
\begin{itemize}
\item[a.] Obtain $\mu_1^*, \sigma_1^{2*}$  by maximizing the log-likelihood $\log f_S$ from $s_i, i=1,\ldots, n$, with starting values given by means of moment estimation of $S$;  plug $\mu_1^*, \sigma_1^{2*}$ into \eqref{meanX0} to estimate the expected position at the time of intervention, i.e. $\hat x=\widehat{\mathbb{E}[X(0)]}$; using $r_i$ and $\hat x$, obtain $\mu_2^*, \sigma_2^{2*}$ as moment estimators for $\mu_2$ and $\sigma_2^2$ when $R|X(0)\sim IG((B-\hat x)/\mu_2, (B-\hat x)^2/\sigma_2^2)$, i.e.
\begin{equation}\label{est0}
\mu_2^*=\frac{B-\hat x}{\bar r}, \qquad \sigma_2^{2*}=\frac{\textrm{emp.var}(R) \mu_2^{3*}}{B-\hat x}
\end{equation}
where $\bar r$ denotes the average of the observations $r_i$. Alternatively, $\mu_2^*$ and $\sigma_2^*$ may be the maximum likelihood estimator \citep{InverseGaussianBook}. Then  $\phi_0=(\mu_1^*,\sigma_1^{2*},\mu_2^*,\sigma_2^{2*})$ is the starting value. When the variances are equal, the starting value is $\phi_0=(\mu_1^*, \sigma_1^{2*},\mu_2^*)$. When the variance is proportional to the mean, obtain $\mu_1^*, k^{*}$  by maximizing the log-likelihood $\log f_S$ from $s_i, i=1,\ldots, n$, with starting values given by means of moment estimation of $S$ through \eqref{specialS};  obtain $\mu_2^*$ as moment estimator for $\mu_2$ from \eqref{specialR}, i.e. $\mu_2^*=(B+k^*)/2 \bar r$. Then set $\phi_0=(\mu_1^*,\mu_2^*,k^*)$.
\end{itemize}
To reduce the influence of the starting value in the optimization procedure, we proceed as follows. Once that $\phi_0$ has been computed, we carry out the estimation procedure, and then we use the obtained estimate $\hat\phi$ as a new starting value $\phi_0$. We repeat this procedure until $\phi_0$ and the estimated parameters yield approximately the same value of $-\log f_{(S,R)}$. 

Often an explicit expression for the inverse of the Fisher information $I(\phi)^{-1}$ is not available, but it can be numerically evaluated. We calculate the $d\times d$ matrix $I(\phi)/n$, for $d=4$ when no assumptions are made and $d=3$ when $\sigma_1^2=\sigma_2^2$ or $\sigma_i=k \mu_i$ using the option {\tt hessian=TRUE} in the {\tt optim} function. Since $I(\phi)$ is symmetric, positive definite square matrix, we invert it by means of its Cholesky decomposition. We first use the function {\tt chol} to compute the Cholesky factorization and then {\tt chol2inv} to invert it.


\end{document}